\documentclass[RNAAS]{aastex62}
\begin{document}

\title{First release of PLATO consortium stellar limb-darkening coefficients}

\correspondingauthor{Giuseppe Morello}
\email{gmorello@iac.es}

\author[0000-0002-4262-5661]{Giuseppe Morello}
\affiliation{Instituto de Astrof\'isica de Canarias (IAC), 38205 La Laguna, Tenerife, Spain}
\affiliation{Departamento de Astrof\'isica, Universidad de La Laguna (ULL), 38206, La Laguna, Tenerife, Spain}
\affiliation{INAF- Palermo Astronomical Observatory, Piazza del Parlamento, 1, 90134 Palermo, Italy}

\author{Jeffrey Gerber}
\affiliation{
Max-Planck Institute for Astronomy, 69117 Heidelberg, Germany}

\author{Bertrand Plez}
\affiliation{LUPM, Univ Montpellier, CNRS, Montpellier, France}

\author{Maria Bergemann}
\affiliation{
Max-Planck Institute for Astronomy, 69117 Heidelberg, Germany}
\affiliation{Niels Bohr International Academy, Niels Bohr Institute, University of Copenhagen, Blegdamsvej 17, DK-2100 Copenhagen, Denmark}

\author{Juan Cabrera}
\affiliation{Institute of Planetary Research, German Aerospace Center, Rutherfordstrasse 2, D-12489 Berlin, Germany}

\author{Hans-G\"unter Ludwig}
\affiliation{Zentrum f\"ur Astronomie der Universit\"at Heidelberg, Landessternwarte, K\"onigstuhl 12, 69117 Heidelberg, Germany}

\author{Thierry Morel}
\affiliation{Space sciences, Technologies and Astrophysics Research (STAR) Institute,
Universit\'e de Li\`ege, Quartier Agora, All\'ee du 6 Ao\^ut 19c, B\^at.
B5c, B4000-Li\`ege, Belgium}


\begin{abstract}
We release the first grid of stellar limb-darkening coefficients (LDCs) and intensity profiles (IPs) computed by the consortium of the PLAnetary Transits and Oscillations of stars (PLATO), the next medium-class (M3) mission under development by the European Space Agency (ESA) to be launched in 2026. We have performed spectral synthesis with \texttt{TurboSpectrum} on a grid of \texttt{MARCS} model atmospheres. Finally, we adopted \texttt{ExoTETHyS} to convolve the high-resolution spectra ($R=2\times10^5$) with the state-of-the-art response functions for all the PLATO cameras, and computed the LDCs that best approximate the convolved IPs. In addition to the PLATO products, we provide new LDCs and IPs for the Kepler mission, based on the same grid of stellar atmospheric models and calculation procedures. The data can be downloaded from the following link: \url{https://doi.org/10.5281/zenodo.7339706}.
\end{abstract}

\section{Introduction}
Stars typically appear brighter in their center than at the edges, due to the so-called limb-darkening effect.
The intensity distribution on the projected stellar disc affects the shape of exoplanetary transit light-curves, since the occulted flux fraction depends on the instantaneous planet's position, in addition to its size \citep{mandel2002,pal2008}. Accurate modeling of stellar limb-darkening is required to recover the correct planet-to-star radius ratio with sub-percent precision  \citep{howarth2011,csizmadia2013,morello2020}. Empirical constraints on the stellar size and limb-darkening effect are also obtained by observations of interferometry \citep{mourard2018,mourard2019} and microlensing events \citep{alcock1997,dominik2004}. Precise measurements of the planetary radii are paramount to constrain their nature, especially for the smaller ones \citep{luque2022}.


PLATO is an ESA M3 mission that is planned for launch in 2026 \citep{rauer2014}. Its main scientific objectives are the detection and bulk characterisation of terrestrial planets in the habitable zone of solar-type stars. PLATO will monitor the photometry of stars over a large field of view in search of planetary transits, following a strategy similar to that of
Kepler \citep{borucki2010} and the Transiting Exoplanet Survey Satellite (TESS, \citealp{ricker2014}).
The payload consists of 24 ``normal'' cameras (N-CAM) with read-out cadence of 25\,s, and two ``fast'' cameras (F-CAM) with read-out cadence of 2.5\,s. The N-CAM will operate in the 0.5-1.0\,$\mu$m wavelength range. The F-CAM have blue and red arms operating at 0.5-0.7\,$\mu$m and 0.65-1.0\,$\mu$m.
In this work, we considered state-of-the-art spectral response functions for the N-CAM, F-CAM blue and red passbands. For the first time, the actual performances obtained from laboratory tests, are taken into account for stellar limb-darkening calculations. Previous unofficial tables of LDCs for PLATO were based on theoretical requirements \citep{kostogryz2022}. Note that ground tests are still ongoing, and in-flight performances will be checked after launch. New grids of LDCs will be released at later stages.

\section{Description of data products}
We adopted 1D line-blanketed hydrostatic LTE MARCS models \citep{gustafsson2008}. We created a randomly distributed grid of 1997 models, with $T_{\rm eff}$ ranging from 4500\,K to 7000\,K, $\log g$ from 3.0 to 5.0, and the solar chemical composition from \citet{magg2022}. Abundances not included in \citet{magg2022} were taken from \citet{asplund2009}.
H, O, Mg, Ca, and Fe were computed following the non-LTE (NLTE) method described in \citet{gerber2022}.
The spectral synthesis calculations were made with the NLTE version of the \texttt{TurboSpectrum} code \citep{plez2012,gerber2022}, with a resolution $R=2\times10^5$ covering the 0.4--1.0\,$\mu$m wavelength range. We computed the intensity spectra for 12 $\mu$-points ($\mu = \cos\theta$, where $\theta$ is the angle between the normal to the surface and the line of sight), distributed according to a Gauss-Radau distribution (GR12). A plane-parallel geometry was assumed for greater computational speed, since sphericity effects are negligible within our grid of models.

Both the sampling of the IP and the optimisation criterion play a crucial role when fitting the LDCs \citep{claret2011,howarth2011,parviainen2015,claret2018,morello2020} Our numerical tests indicate that the chosen GR12 sampling enables robust determination of the LDCs.
We computed and fitted the passband-integrated IPs using the \texttt{ExoTETHyS} package\footnote{\url{https://github.com/ucl-exoplanets/ExoTETHyS}}\citep{morello2020joss,morello2020}, that was specifically optimised for precise modeling of exoplanetary transits down to few parts per million (ppm). In fact, the optimisation algorithm implemented in \texttt{ExoTETHyS} leads to equivalent LDCs to those obtained with the \textit{synthetic-photometry}/\textit{atmosphere-model} method \citep{howarth2011,maxted2018,morello2021rnaas}. We calculated the LDCs for the most popular limb-darkening laws, which are linear \citep{schwarzschild1906}, quadratic \citep{kopal1950}, square-root \citep{diaz-cordoves1992}, power-2 \citep{hestroffer1997}, and claret-4 \citep{claret2000}. We strongly recommend the use of claret-4 coefficients to ensure the most precise approximation of the model IP, although simpler laws could be preferred in some cases to speed up calculations \citep{espinoza2016,morello2017,morello2018}.


\section{Conclusions}
We computed state-of-the-art tables of stellar LDCs and IPs for the PLATO mission, along with new tables for the Kepler mission. For the first time, we adopted the instrumental responses from PLATO laboratory tests. The calculations are based on a grid of \texttt{MARCS} model atmospheres with a fine sampling of the parameter space ranging $T_{\rm eff} = 4500 - 7000\,K$ and $\log{g} = 3.0-5.0$. Technical details, including the accuracy and precision of the data associated with this release note, will be extensively presented in an upcoming article in preparation. The tables are publicly available on Zenodo (\url{https://doi.org/10.5281/zenodo.7339706}).

\acknowledgments
We acknowledge financial support from 
the European Union's Horizon 2020 research and innovation programme under the Marie Sk\l{}odowska-Curie grant agreement No. 895525 and the European Research Council (ERC) grant agreement No. 949173,
the Deutsche Forschungsgemeinschaft (DFG) -- Project-ID 138713538 -- SFB 881,
the Lise Meitner grant from the Max Planck Society,
the French Space Agency (CNES),
and the Belspo contracts for PLATO mission development.

\bibliography{mybib.bib}

\end{document}